\documentclass[12pt]{article}
\usepackage{psfig}
\newcommand{\gsim}{\lower.7ex\hbox{$
    \;\stackrel{\textstyle>}{\sim}\;$}}
\newcommand{\lsim}{\lower.7ex\hbox{$
    \;\stackrel{\textstyle<}{\sim}\;$}}
\def\order#1{{\cal O}\left(#1\right)}
\def\eq#1{(\ref{#1})}

\newcommand{\ba}{\begin{eqnarray}}
\newcommand{\ea}{\end{eqnarray}}
\def\dd{{\rm d}}

\begin{document}

\setlength{\unitlength}{1mm}

\thispagestyle{empty}


\begin{flushright}
{\bf Alberta Thy 03-01}\\
{\bf BNL-HET-01/4}\\
{\bf hep-ph/0102122}
\end{flushright}
\vspace{0.6cm}
\boldmath
\begin{center}
\Large\bf\boldmath The Muon Anomalous Magnetic Moment:\\ A Harbinger For
``New Physics''
\end{center}
\unboldmath
\vspace{0.8cm}

\begin{center}

{\large Andrzej Czarnecki}\\[2mm]
{\sl Department of Physics, University of Alberta\\
Edmonton, AB, Canada T6G 2J1}

\vspace*{7mm}
{\large William J. Marciano}\\[2mm]
{\sl Physics Department, Brookhaven National Laboratory,\\
Upton, NY 11973, USA}

\end{center}

\vfill

\begin{abstract}
QED, Hadronic, and Electroweak Standard Model contributions to the
muon anomalous magnetic moment, $a_\mu\equiv (g_\mu-2)/2$, and their
theoretical uncertainties are scrutinized.  The status and
implications of the recently reported 2.6 sigma experiment vs.~theory
deviation $a_\mu^{\rm exp}-a_\mu^{\rm SM} = 426(165)\times 10^{-11}$
are discussed.  Possible explanations due to supersymmetric loop
effects with $m_{\rm SUSY} \simeq 55 \sqrt{\tan\beta}$ GeV, radiative
mass mechanisms at the 1--2 TeV scale and other ``New Physics''
scenarios are examined.
\end{abstract}
\vfill

\newpage

\section{Introduction}
\label{sec1}
Leptonic anomalous magnetic moments have traditionally  provided 
precision tests of the
Standard Model (SM) and stringent constraints on potential ``New Physics''
effects.  In the case of the electron, comparing the extraordinary
measurements of $a_e \equiv (g_e-2)/2$ at the University of Washington
\cite{Dehmelt87} 
\ba
a_{e^-}^{\rm exp} &=& 0.001\, 159\, 652 \,   188\, 4(43),
\nonumber \\
a_{e^+}^{\rm exp} &=& 0.001\, 159\, 652 \,   187\, 9(43),
\ea
with the prediction
\cite{Mohr99,Laporta:1997zy,Hughes:1999fp,Czarnecki:1998nd} 
\ba
a_e^{\rm SM} &=& {\alpha\over 2\pi} 
 -0.328\,478\,444\,00 \left( {\alpha\over \pi}\right)^2
 +1.181\,234\,017 \left( {\alpha\over \pi}\right)^3
\nonumber\\
&& \hspace*{-9mm}
 -1.5098(384)\left( {\alpha\over \pi}\right)^4
+1.66(3)\times 10^{-12} 
\mbox{(hadronic \& electroweak loops)}
\ea
currently
provides the best determination of the fine structure constant
\cite{Kinoshita:1996vz}, 
\ba
\alpha^{-1}(a_e) = 137.035\,999\,58(52).
\label{eq3}
\ea
To test the Standard Model requires comparison with 
an alternative measurement of
$\alpha$ with comparable accuracy.  Unfortunately, the next best
determination of $\alpha$, from the quantum Hall effect \cite{Mohr99},
\ba
\alpha^{-1}(qH) = 137.036\,003\,00(270),
\ea
has a considerably larger error.   If one assumes that $\left| 
\Delta a_e^{\rm
New\ Physics}\right| \simeq m_e^2 / \Lambda^2$, where $\Lambda$
approximates the 
scale of ``New Physics'', then the agreement between
$\alpha^{-1}(a_e)$ and $\alpha^{-1}(qH)$ currently probes $\Lambda
\lsim \order{\mbox{100 GeV}}$.  
To access the much more interesting  $\Lambda
\sim \order{\mbox{TeV}}$ region would require an order of magnitude
improvement in $a_{e}^{\rm exp}$ (technically feasible \cite{Gab94}),
an improved calculation of the 4-loop QED contribution to $a_e^{\rm
SM}$ and a much better independent measurement of $\alpha^{-1}$ by
almost two orders of magnitude.  The last requirement, although
extremely challenging, is perhaps most
likely to come \cite{Kinoshita:1996vz}  from combining the already
precisely measured Rydberg constant with  a much better determination
of $m_e$.  

We should note that for potential ``New Physics'' (NP) effects linear
in 
the electron mass, $\Delta a_e^{\rm NP} \sim m_e/\Lambda$, naively,
one is currently probing a much more impressive $\Lambda \sim
\order{10^7\mbox{ GeV}}$ and the possible advances described above
would explore $\order{10^9\mbox{ GeV}}$! However, we subsequently
argue that such linear ``New Physics'' effects are generally
misleading because the associated physics is likely to also give
unacceptably large corrections to the electron mass.

Improvements in the measurement of the muon's anomalous magnetic
moment have also been impressive.  A series of dedicated experiments
at CERN that ended in 1977 found
\cite{PDG98} 
\ba
a_\mu^{\rm exp} = 116 \, 592 \, 300 (840) \times 10^{-11} \qquad
\mbox{(CERN 1977)}.
\label{eq5}
\ea
More recently, an ongoing experiment (E821) at Brookhaven National
Laboratory has been running with much higher statistics and a very
stable, well measured
magnetic field in its storage ring.  Based on $\mu^+$ data taken
through  1998, combined with the earlier CERN result in (\ref{eq5}),
it reported \cite{Brown:2000sj}
\ba
a_\mu^{\rm exp} = 116 \, 592 \, 050 (460) \times 10^{-11} \qquad
\mbox{(CERN'77 + BNL'98)}.
\label{eq6}
\ea
That group has just announced a much higher statistics result based on
1999 data \cite{Brown2001},
\ba
a_\mu^{\rm exp} = 116\,592\,020 (160) \times 10^{-11}
\quad \mbox{(BNL'99)}.
\label{eq7m}
\ea
Their finding is very consistent with Eq.~\eq{eq6}.  When simply
averaged together, we find
\ba
a_\mu^{\rm exp}(\rm Average) = 116\,592\,023 (151) \times 10^{-11}
\quad \mbox{(CERN'77 + BNL'98 \& '99)}.
\label{eq7}
\ea
The ultimate goal of the experiment (which has its final scheduled run
with $\mu^-$
during 2001) is $\pm 40\times 10^{-11}$, about a factor of 20
improvement relative to the classic CERN experiments and a factor of
3.5 better than the average in Eq.~(\ref{eq7}).  Even the inclusion of
already existing data from the 2000 run is expected to reduce the
error in  Eq.~(\ref{eq7}) by more than  a factor of 2 within the
coming year. 

Although $a_\mu^{\rm exp}$ is currently about 350 times less precise
than $a_e^{\rm exp}$, it is much more sensitive to hadronic and
electroweak quantum loops as well as ``New Physics'' effects, since
such contributions \cite{km90} are generally proportional to $m_l^2$.
The $m_\mu^2/m_e^2 \simeq 40\,000$ enhancement more than compensates
for the reduced experimental precision and makes $a_\mu^{\rm exp}$ a
much better probe of short-distance phenomena.  Indeed, as we later
illustrate, a deviation in $a_\mu^{\rm exp}$ from the Standard Model
prediction, $a_\mu^{\rm SM}$, even at its current level of sensitivity
can quite naturally be interpreted as
the appearance of ``New Physics'' such as supersymmetry at 100-450
GeV, or other even higher scale phenomena, exciting prospects.  Of
course, before making such an interpretation, one must have a reliable
theoretical prediction for $a_\mu^{\rm SM}$ with which to compare, an
issue that we address in the next section.

Before leaving the comparison between $a_e^{\rm exp}$ and $a_\mu^{\rm
exp}$, we should remark that for cases where ``New Physics''
contributions to $a_l$ scale as $m_l/\Lambda$, roughly equal
sensitivity in $\Lambda$ ($\sim 10^{7} \mbox{ GeV}$) currently exists
for both types of measurements.  However, as previously mentioned,
such effects are in our view artificial.

\section{Standard Model Prediction For $a_\mu$}
\subsection{QED Contribution}
The QED contribution to $a_\mu$ has been computed (or estimated) 
through 5 loops
\cite{Czarnecki:1998nd,Mohr99} 
\ba
a_\mu^{\rm QED} 
 &=& {\alpha\over 2\pi} 
 +0.765\,857\,376(27) \left( {\alpha\over \pi}\right)^2
 +24.050\,508\,98(44) \left( {\alpha\over \pi}\right)^3
\nonumber\\
&& +126.07(41)\left( {\alpha\over \pi}\right)^4 
+930(170)\left({\alpha\over \pi}\right)^5. 
\ea
Growing coefficients in the $\alpha/\pi$ expansion reflect the
presence of large $\ln{m_\mu\over m_e}\simeq 5.3$ terms coming from
electron loops.  Employing the value of $\alpha$ from $a_e$ in
eq.~(\ref{eq3}) leads to 
\ba
a_\mu^{\rm QED} = 116\,584\,705.7(2.9)\times 10^{-11}.
\label{eq8}
\ea 
The current uncertainty is well below the $\pm 40\times 10^{-11}$
ultimate experimental error anticipated from E821 and should,
therefore, play no essential role in the confrontation between theory
and experiment.

\subsection{Hadronic Loop Corrections}

\begin{figure}[thb]
\hspace*{45mm}
\psfig{figure=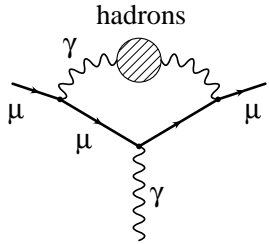,width=35mm}
\caption{Leading hadronic vacuum polarization corrections to $a_\mu$.}
\label{fig:had}
\end{figure}

Starting at $\order{\alpha^2}$, hadronic loop effects contribute to
$a_\mu$ via vacuum polarization (see Fig.~\ref{fig:had}).  A first
principles QCD calculation 
of that effect does not exist.  Fortunately, it is possible to
evaluate the leading effect via the dispersion integral \cite{GRaf}
\ba
a_\mu^{\rm Had} (\mbox{vac. pol.}) = {1\over 4\pi^3}
\int_{4m_\pi^2}^\infty {\dd s} \, 
 K\left(s\right) \, \sigma^0(s)_{e^+e^- \to {\rm hadrons}},
\label{eq9}
\ea
where 
$\sigma^0(s)_{e^+e^- \to {\rm hadrons}}$ means QED vacuum
polarization and some other extraneous radiative corrections
(e.g. initial state radiation) have been
subtracted from measured $e^+e^-\to$ hadrons cross sections, and 
\ba
K(s) &=& x^2\left(1-{x^2\over 2}\right)
 + (1+x)^2 \left( 1+{1\over x^2}\right)
 \left[ \ln(1+x) -x+{x^2\over 2}\right]
 +{1+x\over 1-x}\, x^2 \ln x
\nonumber \\
x&=& {1-\sqrt{1-4m_\mu^2/s} \over 1+\sqrt{1-4m_\mu^2/s}}.
\ea
Detailed studies of eq.~\eq{eq9} have been carried out by a number of
authors \cite{Alemany:1997tn,Davier:1998si,Davier:1999xy,Davier:1998iz,%
Jeg95,kinoshita85,light,Erler:2000nx}. The most precise published
analysis to date, due to Davier
and H\"ocker \cite{Davier:1998si,Davier:1999xy,Davier:1998iz},
found
\ba
a_\mu^{\rm Had}(\mbox{vac. pol.}) = 6924(62)\times 10^{-11}.
\label{eq10}
\ea
It employed experimental $e^+e^-$ data, hadronic tau decays,
perturbative QCD and sum rules to minimize the uncertainty in that
result. The  contributions coming from various energy regions are
illustrated in Table \ref{tab1}.
\begin{table}[htb]
\caption{Contributions to $a_\mu^{\rm Had}(\mbox{vac. pol.})$ from
different energy regions as found by Davier and H\"ocker
\protect\cite{Davier:1998si,Davier:1999xy,Davier:1998iz}.}
\label{tab1} 
\vspace*{3mm}
\begin{center}
\begin{tabular}{l@{\hspace{10mm}}r}
\hline
\hline
\\[-4mm]
$\sqrt{s}$ (GeV)  & $a_\mu^{\rm Had}(\mbox{vac. pol.})\times 10^{11}$
\\[.8mm]
\hline
$2m_\pi - 1.8$  & $6343\pm 60$ \\
$1.8 - 3.7$  & $338.7\pm 4.6$ \\
$3.7 - 5+\psi(1S,2S)$  & $143.1\pm 5.4$ \\
$5-9.3$  & $68.7\pm 1.1$ \\
$9.3-12$  & $12.1\pm 0.5$ \\
$12-\infty$  & $18.0\pm 0.1$ \\
\hline
\\[-4mm]
Total    & $6924\pm 62$\\
\hline
\hline
\end{tabular}
\end{center}
\end{table}

It is clear from Table \ref{tab1} that the final result and its
uncertainty are dominated by the low energy region.  In fact, the
$\rho(770 \mbox{ MeV})$ resonance provides about 72\% of the total
hadronic contribution to $a_\mu^{\rm Had}(\mbox{vac. pol.})$.

To reduce the uncertainty in the $\rho$ resonance region, Davier and
H\"ocker employed $\Gamma(\tau\to \nu_\tau \pi^- \pi^0)/\Gamma(\tau
\to \nu_\tau \bar \nu_e e^-)$ data to supplement $e^+e^-\to
\pi^+\pi^-$ cross-sections.  In the $I=1$ channel they are related by
isospin.  Currently, tau decay data is experimentally more precise and
in principle has the advantage of being self-normalizing if both
$\tau\to\nu_\tau \pi^-\pi^0$ and $\tau\to\nu_\tau \overline{\nu}_e e$
are both measured in the same experiment.

An issue in the use of tau decay data is the magnitude of isospin
violating corrections due to QED and the $m_d-m_u$ mass difference.  A
short-distance QED correction \cite{Marciano:1988vm}  of about $-2\%$
was applied to the hadronic tau decay data and isospin violating
effects such as 
$m_{\pi^\pm}-m_{\pi^0}$  phase space and $\rho^\pm-\rho^0$ differences 
have been accounted
for.  Other uncorrected differences are estimated to be about
$\pm 0.5\%$ and included in the hadronic uncertainty.  

Although the $\pm 0.5\%$ error assigned to the use of tau decay data
appears reasonable, it has been questioned
\cite{EidelmanPriv,Jegerlehner:1999hg}.  More recent preliminary
$e^+e^-\to \pi^+\pi^-$ data from Novosibirsk \cite{EidelmanPriv} seems
to suggest a potential difference with corrected hadronic tau decays
which could compromise the estimated $a_\mu^{\rm Had}$ in
Eq.~(\ref{eq10}).  It is not clear at this time whether the difference
is due to additional isospin violating corrections to hadronic tau
decays, normalization issues 
\cite{Anderson:1999ui}, or radiative corrections to $e^+e^-\to $
hadrons data which must be accounted for in any precise comparison
\cite{Marciano:1992pr}.  Resolution of this issue is extremely
important.

A more conservative approach might be to ignore the tau data and use QCD
theory input as little as possible.  In an (unpublished) update of earlier
work \cite{Jeg95}, Jegerlehner found from such an approach
\ba
a_\mu^{\rm Had}(\mbox{vac. pol.}) = 6988(111) \times 10^{-11}
\quad (\mbox{Jegerlehner 2000, preliminary \cite{Jeg2000}}).
\label{eq12n}
\ea
Within their quoted errors, Eqs.~(\ref{eq10}) and (\ref{eq12n}) agree
but the central values differ by $64\times 10^{-11}$.  The sign of the
difference between Eq.~\eq{eq12n} and Eq.~\eq{eq10} may be a little
misleading, since tau data tends to favor a larger contribution to
$a_\mu^{\rm Had}$ from the $\rho$ than $e^+e^-\to$ hadrons
\cite{Anderson:1999ui}. Anticipated new
results for $e^+e^-\to $ hadrons at Novosibirsk should reduce the
uncertainty in (\ref{eq12n}) by nearly a factor of 2 without requiring
tau data.  It will be interesting to see what happens to its central
value.

Evaluation of the 3-loop hadronic vacuum polarization contribution to
$a_\mu$  has been updated to \cite{Krause:1997rf,kinoshita85} 
\ba
\Delta a_\mu^{\rm Had}(\mbox{vac. pol.}) = -100 (6) \times 10^{-11}.
\label{eq11}
\ea
Light-by-light hadronic diagrams have been evaluated using chiral
perturbation theory.  An average
\cite{Davier:1998si,Davier:1999xy,Davier:1998iz} of two recent studies
\cite{Bijnens:1996xf,Hayakawa:1998rq} gives
\ba
\Delta a_\mu^{\rm Had}(\mbox{light-by-light}) = -85(25)\times
10^{-11}.
\label{eq12}
\ea
Adding those contributions to Eqs.~(\ref{eq10})
leads to the total hadronic contribution
\ba
a_\mu^{\rm Had} = 6739(67)\times 10^{-11} \quad\mbox{(Davier \&
H\"ocker, 1998)}
\label{eq13}
\ea
which we will subsequently use in comparison of theory and
experiment.  However, we note that a more conservative approach might
employ a larger uncertainty such as found using Jegerlehner's
unpublished result  in Eq.~(\ref{eq12n}),
\ba
a_\mu^{\rm Had} = 6803(114)\times 10^{-11}\quad\mbox{(Jegerlehner
2000, unpublished)}.
\label{eq13alt}
\ea
At the very least, one should be mindful of the difference between the
two and the need to further justify the use of tau decay data and
low-energy perturbative QCD.  
The uncertainties in those results represent the main theoretical error
in $a_\mu^{\rm SM}$.  It would be very valuable to supplement the
above evaluation of $a_\mu^{\rm Had}$ with lattice calculations (for
the light-by-light  contribution) and further 
improved $e^+e^-$ data (beyond ongoing experiments).  An ultimate goal of
$\pm 40 \times 10^{-11}$ or smaller appears to be within reach and is
well matched to the prospectus of experiment E821 at Brookhaven which
aims for a similar level of accuracy.

\subsection{Electroweak corrections}

\begin{figure}[thb]
\hspace*{-8mm}
\begin{minipage}{16.cm}
\vspace*{6mm}
\[
\mbox{ 
\begin{tabular}{ccc}
\psfig{figure=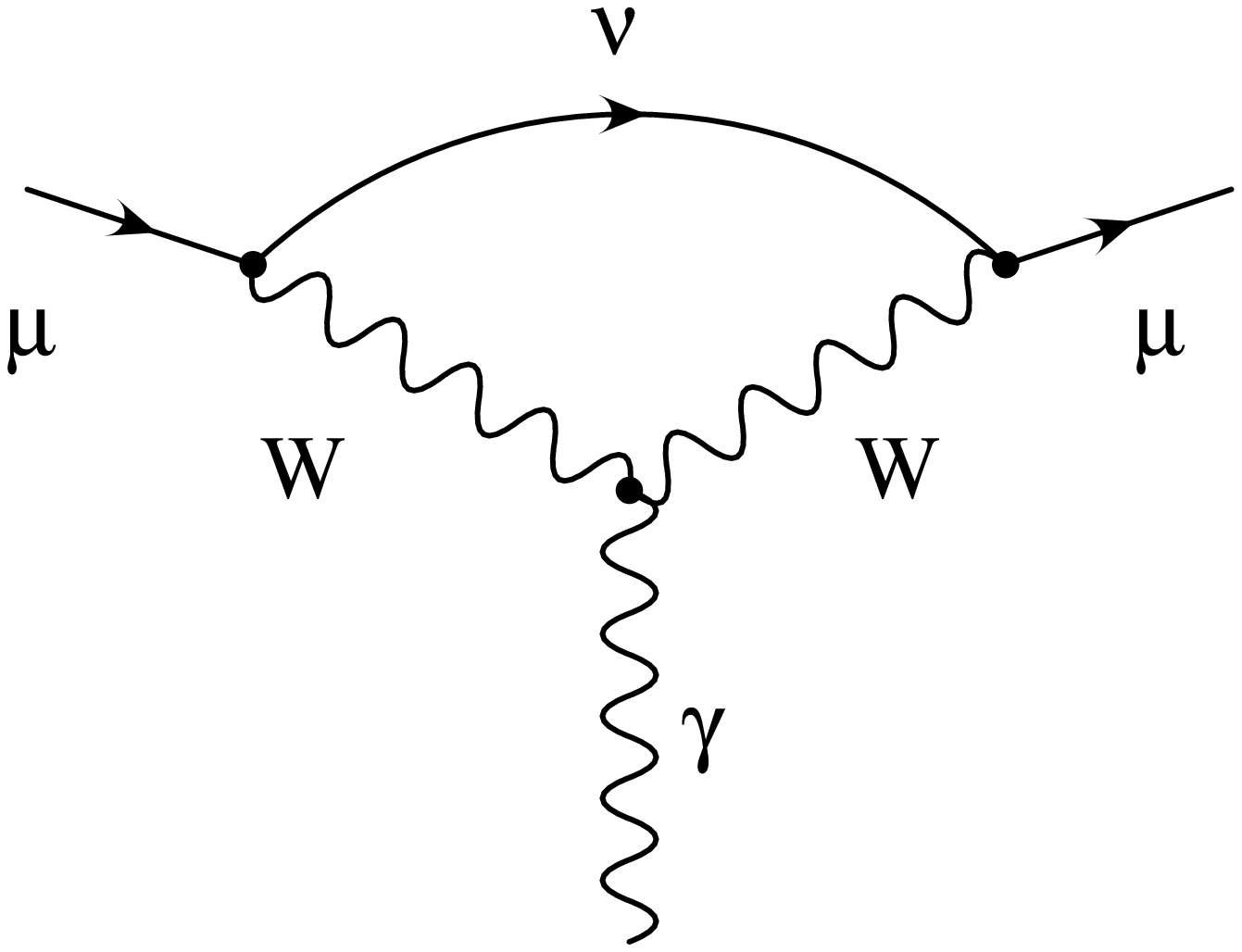,width=40mm} 
& \hspace*{6mm}
\psfig{figure=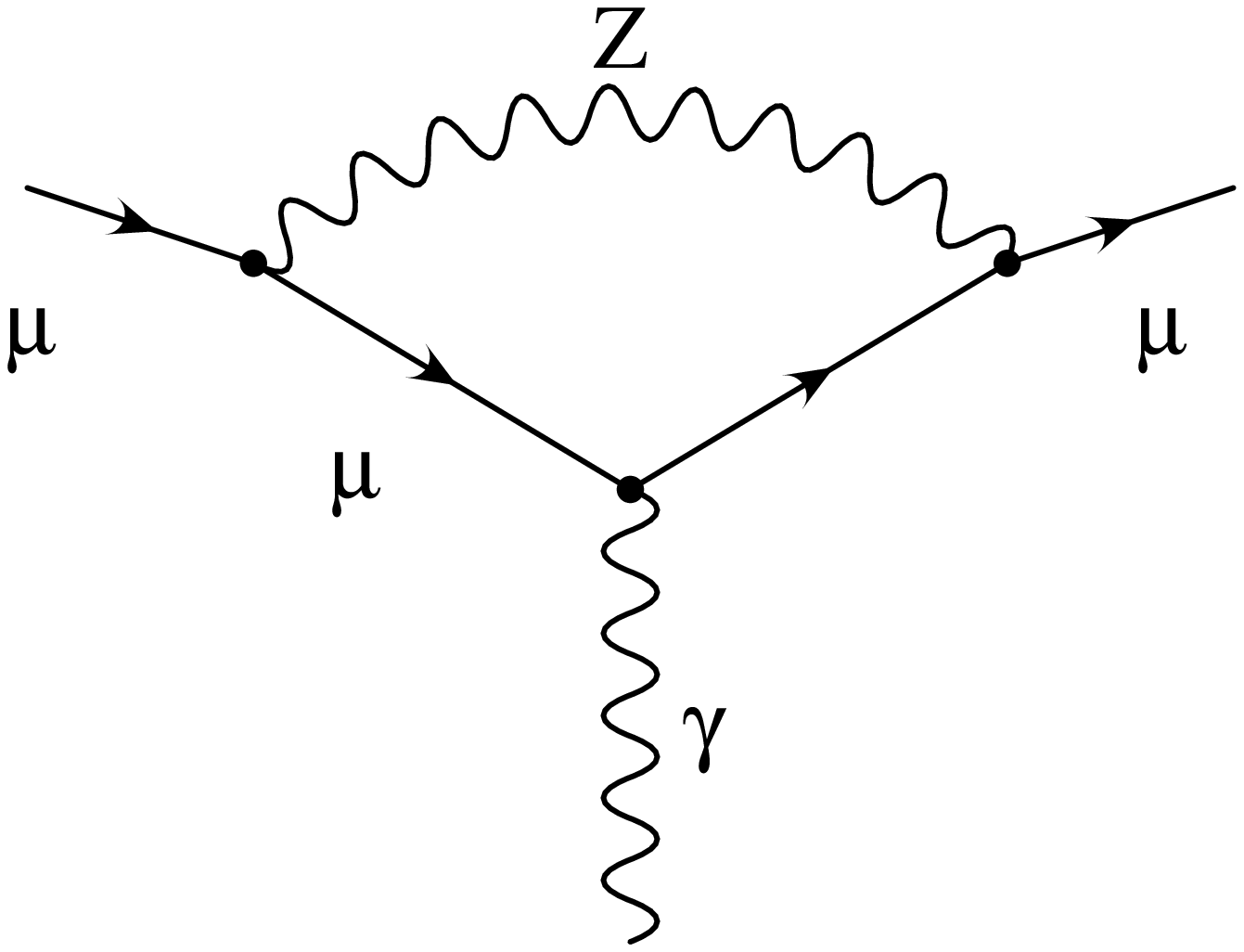,width=40mm} 
& \hspace*{6mm}
\psfig{figure=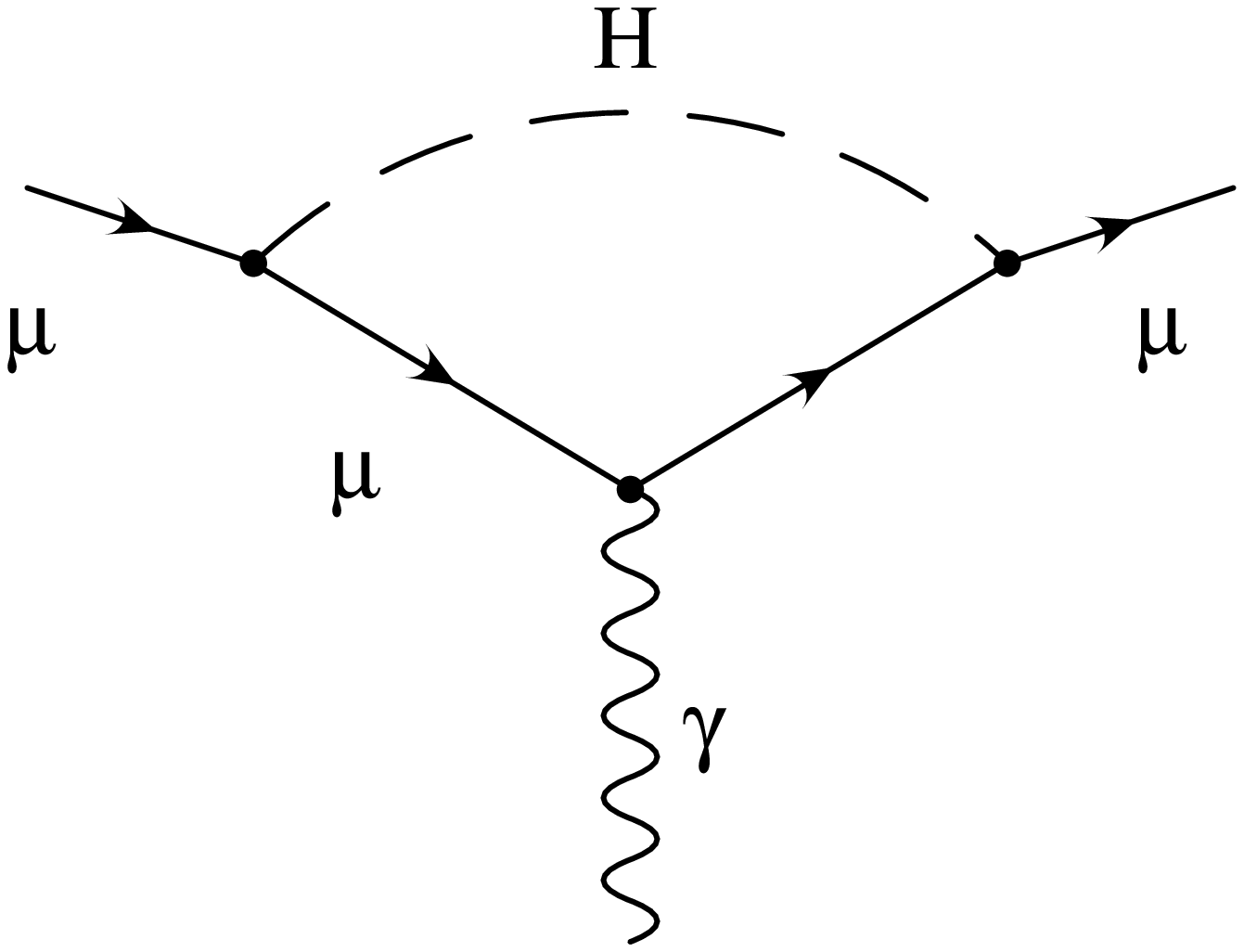,width=40mm} 
\\[2mm]
(a) &\hspace*{6mm} (b)&\hspace*{6mm} (c)
\end{tabular}
}
\]
\end{minipage}
\caption{One-loop electroweak radiative corrections to $a_\mu$.}
\label{fig:oneloop}
\end{figure}
The one-loop electroweak radiative corrections to $a_\mu$ (see
Fig.~\ref{fig:oneloop})  are
predicted in the Standard Model to be 
\cite{Brodsky:1967mv,Burnett67,Jackiw72,fls72,Bars72,ACM72,Bardeen72}
\begin{eqnarray}
\lefteqn{a_\mu^{\rm EW}(\rm 1\,loop) =
{5\over 3}{G_\mu m_\mu^2\over 8\sqrt{2}\pi^2}}
\nonumber\\ && \times
\left[1+{1\over 5}(1-4\sin^2\theta_W)^2
+ {\cal O}\left({m_\mu^2 \over M^2}\right) \right]
\nonumber \\
&& \approx 195 \times 10^{-11}
\label{eq14}
\end{eqnarray}
where $G_\mu = 1.16637(1) \times 10^{-5}$ GeV$^{-2}$,
$\sin^2\theta_W\equiv 1-M_W^2/M_Z^2\simeq 0.223$. 
and $M=M_W$ or
$M_{\rm Higgs}$.  The original goal of E821 at Brookhaven was to
measure that predicted effect at about the 5 sigma level (assuming
further reduction in the hadronic uncertainty).  Subsequently, it was
pointed out \cite{KKSS} that two-loop electroweak contributions are
relatively large due to the presence of $\ln m_Z^2/m_\mu^2\simeq 13.5$
terms.  A full two-loop calculation \cite{CKM96,CKM95}, including
low-energy hadronic electroweak loops  \cite{Peris:1995bb,CKM95},
found for 
$m_H \simeq 150$ GeV (with little sensitivity to the exact value)
\ba
a_\mu^{\rm EW}(\mbox{2 loop}) = -43(4) \times 10^{-11},
\label{eq15}
\ea
where the quoted error is a conservative estimate of hadronic, Higgs,
and higher-order corrections.  Combining eqs.~\eq{eq14} and \eq{eq15}
gives the electroweak contribution
\ba
a_\mu^{\rm EW} = 152(4)\times 10^{-11}.
\label{eq16}
\ea
Higher-order leading logs of the form $(\alpha\ln m_Z^2/m_\mu^2)^n$,
$n=2,3,\ldots$ can be computed via renormalization group techniques
\cite{Degrassi:1998es}.  Due to cancellations between the running of
$\alpha$ and anomalous dimension effects, they give a relatively small
$+0.5\times 10^{-11}$ contribution to $a_\mu^{\rm EW}$.  It is safely
included in the uncertainty of eq.~\eq{eq16}.

\subsection{Comparison with Experiment}
The complete Standard Model prediction for $a_\mu$ is 
\ba
a_\mu^{\rm SM}= a_\mu^{\rm QED} + a_\mu^{\rm Had} + a_\mu^{\rm EW}.
\ea
Combining eqs.~\eq{eq8}, \eq{eq13} and \eq{eq16}, one finds
\ba
a_\mu^{\rm SM}= 116\,591\,597(67) \times 10^{-11},
\label{eq22n}
\ea
or, using  eqs.~\eq{eq8}, along with 
the more conservative \eq{eq13alt} and \eq{eq16},
$a_\mu^{\rm SM}= 116\,591\,661(114) \times 10^{-11}$.
Comparing Eq.~(\ref{eq22n}) with the current experimental average in
Eq.~\eq{eq7} gives
\ba
a_\mu^{\rm exp}-a_\mu^{\rm SM}= 426\pm 165\times 10^{-11}.
\label{eq19}
\ea
The roughly $2.6\sigma$ difference is very exciting.  It may be an
indicator or harbinger of
contributions from ``New
Physics'' beyond the Standard Model.  At 
90\% CL, one finds
\ba
215 \times 10^{-11}\le a_\mu({\rm New\ Physics}) \le 637 \times 10^{-11},
\label{eq20}
\ea
which suggests a relatively large ``New Physics'' effect, even larger
than the predicted electroweak contribution, is starting to be seen.
As we show in the next section, several realistic examples of ``New
Physics'' could quite easily lead to $a_\mu({\rm New\ Physics})\sim
\order{426\times 10^{-11}}$ and might be responsible for the apparent
deviation. If that is the case, the difference in Eq.~\eq{eq19} should
increase to a 6 or more sigma effect as E821 is completed and the
hadronic uncertainties in $a_\mu^{\rm SM}$ are further reduced.

\section{``New Physics'' effects}

Since the anomalous magnetic moment comes from a dimension 5
operator, ``New Physics'' (i.e.~beyond the Standard Model 
expectations) will contribute to $a_\mu$ via induced 
quantum loop effects
(rather than tree level).
Whenever a new model or Standard Model extension is proposed, such
effects are examined and
$a_\mu^{\rm exp}-a_\mu^{\rm SM}$ is often employed to constrain or
rule it out.

In this section we describe several  examples of interesting
``New Physics'' probed by $a_\mu^{\rm exp}-a_\mu^{\rm SM}$. Rather
than attempting to be inclusive, we concentrate on two general
scenarios: 1) Supersymmetric loop effects which can be substantial and
would be heralded as the most likely explanation if the deviation in
$a_\mu^{\rm exp}$ is confirmed and 2) Models of radiative muon mass
generation which predict $a_\mu({\rm New\ Physics}) \sim m_\mu^2 /
M^2$ where $M$ is the scale of ``New Physics''. 
Either case is capable of explaining the apparent deviation in
$a_\mu^{\rm exp}-a_\mu^{\rm SM}$ exhibited in Eq.~(\ref{eq19}).
 Other examples of
potential ``New Physics'' contributions to $a_\mu$ are only briefly
discussed.

\subsection{Supersymmetry}

\begin{figure}[thb]
\hspace*{-18mm}
\begin{minipage}{16.cm}
\vspace*{6mm}
\[
\mbox{ 
\begin{tabular}{cc}
\psfig{figure=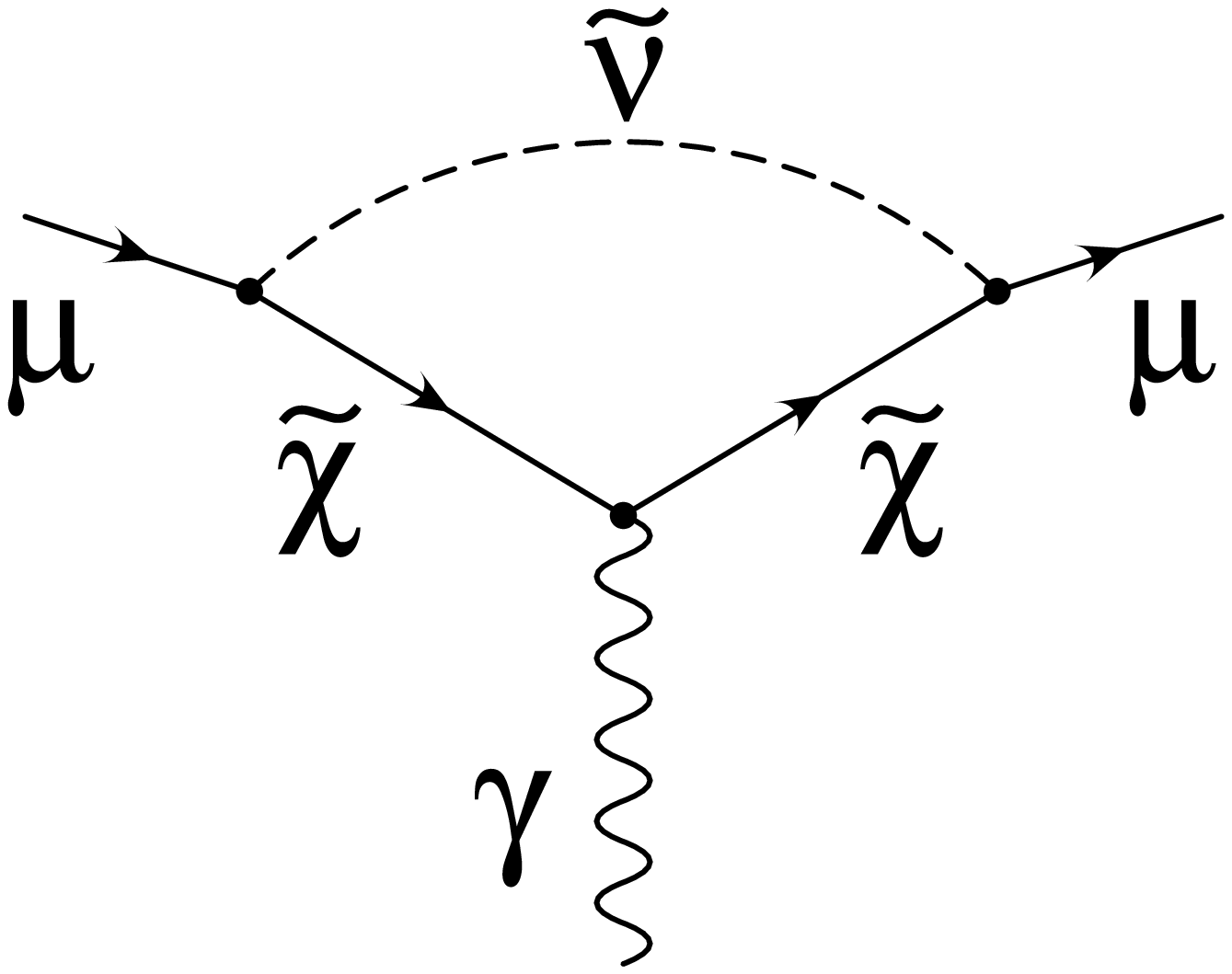,width=35mm,bbllx=72pt,bblly=291pt,%
bburx=544pt,bbury=540pt} 
& \hspace*{10mm}
\psfig{figure=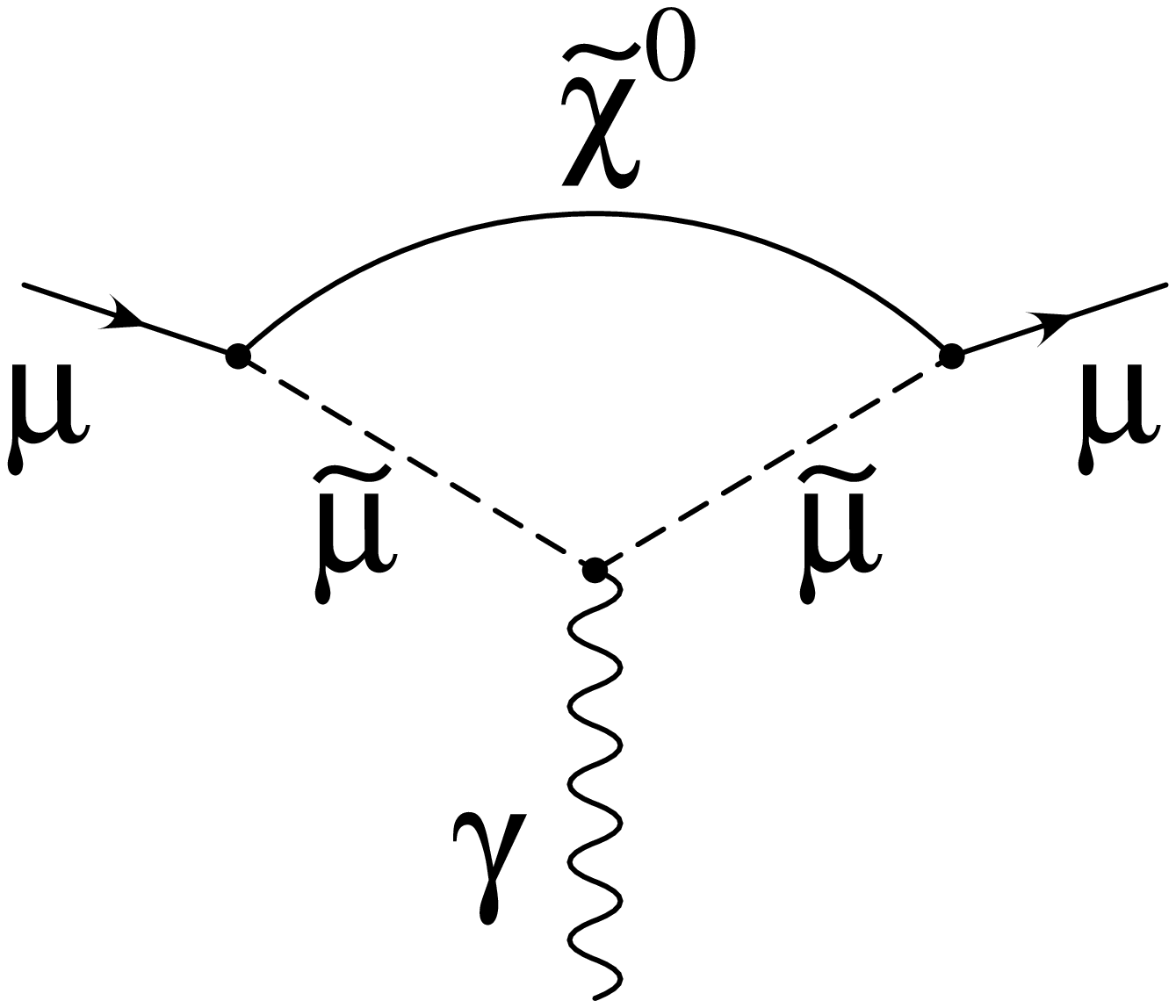,width=35mm,bbllx=72pt,bblly=291pt,%
bburx=544pt,bbury=540pt} 
\\[2mm]
(a) &\hspace*{10mm} (b)
\end{tabular}
}
\]
\end{minipage}
\caption{Supersymmetric loops contributing to the muon anomalous
magnetic moment.}
\label{fig1}
\end{figure}

The supersymmetric contributions to $a_\mu$ stem from
smuon--neutralino and sneutrino-chargino loops (see Fig.~\ref{fig1}).
They include 2 chargino and 4 neutralino states and could in
principle entail slepton mixing and phases.  Depending on SUSY masses,
mixing and other parameters, the contribution of $a_\mu^{\rm SUSY}$
can span a broad range of possibilities.  Studies have been carried
out for a variety of models where the parameters are specified.  
Here we give a generic discussion primarily
intended to illustrate the strong likelihood that evidence for
supersymmetry can be
inferred from  $a_\mu^{\rm exp}$ and may in fact be the natural
explanation for the apparent deviation from SM theory reported by
E821. 

Early studies of the supersymmetric contributions $a_\mu^{\rm SUSY}$
were carried out in the context of the minimal SUSY standard model
(MSSM)
\cite{fayet80,Grifols:1982vx,Ellis:1982by,Barbieri:1982aj,Romao:1985pn,%
Kosower:1983yw,Yuan:1984ww,Vendramin:1989rd}, in an $E_6$
string-inspired model \cite{Grifols:1986vr,Morris:1988fm}, and in an
extension of the MSSM with an additional singlet
\cite{Frank:1988yn,Francis:1991pi}.  An important observation was made
in \cite{Lopez:1994vi}, namely that some of the contributions are
enhanced by the ratio of Higgs' vacuum expectation values,
$\tan\beta\equiv \langle \Phi_2\rangle/\langle \Phi_1\rangle$, which
in some models is large (in some cases of order $m_t/m_b\approx 
40$).  In addition, larger values of $\tan\beta \gsim 2$ are generally
in better accord with the recent 
LEP II Higgs mass bound $m_H\gsim 113 $
GeV and, therefore, currently favored.  
The main contribution is generally due to the
chargino-sneutrino diagram (Fig.~\ref{fig1}a), which is enhanced by a
Yukawa coupling in the muon-sneutrino-Higgsino vertex (charginos are
admixtures of Winos and Higgsinos).

The leading effect from Fig.~\ref{fig1}a
is approximately given in the large $\tan\beta$
limit by 
\ba
\left|a_\mu^{\rm SUSY}\right|
 \simeq {\alpha(M_Z)\over 8\pi\sin^2\theta_W}\,
{m_\mu^2\over \widetilde{m}^2}\,\tan\beta \left( 1-{4\alpha\over
\pi}\ln {\widetilde m\over m_\mu}\right),
\ea
where $\widetilde{m}=m_{\rm SUSY}$ 
represents a typical SUSY loop mass.  (Chargino-
and sneutrino-masses are actually assumed degenerate in that
expression \cite{Moroi:1996yh}; otherwise, $\widetilde{m}$ is
approximately the
heavier mass scale.)  Also, we have included a 7--8\% suppression
factor due to leading 2-loop EW effects. Like most ``New Physics'' effects,
SUSY loops contribute directly to the dimension 5 magnetic dipole
operator.  In that case, they are subject to the same EW suppression
factor as the $W$ loop contribution to $a_\mu^{\rm EW}$.  From the
calculation in Ref.~\cite{CKM96,Degrassi:1998es}, one finds a leading
log suppression factor 
\ba 
1-{4\alpha \over \pi} \ln {M\over m_\mu}
\label{eq:sup}
\ea
where $M$ is the characteristic ``New Physics'' scale.  For
$M\sim 200$ GeV, that factor corresponds to about a 7\% reduction. 

Numerically, one expects in the large $\tan\beta$ regime (after a
small negative contribution from Fig.~\ref{fig1}b is included, again
assuming degenerate masses)
\ba
\left|a_\mu^{\rm SUSY}\right|
 \simeq 130\times 10^{-11} \left( 100 {\rm\ GeV}\over 
\widetilde{m}\right)^2 \tan\beta,
\label{eq22}
\ea
where $a_\mu^{\rm SUSY}$ generally has the same sign as the
$\mu$-parameter in SUSY models.

Ref.~\cite{Lopez:1994vi} found that E821 will be a stringent test of a
class of supergravity models.  However, in the minimal SU(5) SUGRA
model, $\tan\beta$ is already severely constrained by the proton decay
lifetime and no significant $a_\mu^{\rm SUSY}$ is possible.  Extended
models, notably SU(5)$\times$U(1) escape that bound and can induce
large effects.

Supersymmetric effects in $a_\mu$ were subsequently computed in a
variety of models.  Constraints on MSSM were examined in
\cite{Moroi:1996yh,Cho:2000sf}.  MSSM with large CP-violating phases
was studied in \cite{Ibrahim:1999aj}.  Ref.~\cite{Brignole:1999gf}
examined models with a superlight gravitino.  Detailed studies of
$a_\mu^{\rm SUSY}$ were carried out in models constrained by various
assumptions on the SUSY-breaking mechanism: gauge-mediated
\cite{Carena:1997qa,Mahanthappa:1999ta}, 
SUGRA \cite{Nath95,Goto:1999mk,Blazek:1999hb}, and anomaly-mediated
\cite{Chattopadhyay:2000ws}. 

Rather than focusing on a specific model, 
we simply employ for illustration the large $\tan\beta$ approximate
formula in eq.~\eq{eq22} with degenerate SUSY masses 
and the current constraint in eq.~\eq{eq19}.
Then we find (for positive sgn$(\mu)$) from comparison with
Eq.~\eq{eq19} 
\ba
\tan\beta \left( {100 \mbox{ GeV}\over \widetilde m}\right)^2 \simeq
3.3 \pm 1.3 ,
\label{eq23}
\ea
or 
\ba
 \widetilde m \simeq (55\mbox{ GeV})\sqrt{\tan\beta}.
\label{eq30n}
\ea
 (Of course, in specific models with
non-degenerate gauginos and sleptons, a more detailed analysis is
required, but here we only want to illustrate roughly the scale of
supersymmetry being probed.)
Negative $\mu$ models give the opposite sign contribution to $a_\mu$ and 
are strongly disfavored.  

For large $\tan\beta$ in the range $4\sim 40$, where the approximate
results given above should be valid, one finds (assuming $\widetilde m
> 100$ GeV from other experimental constraints)
\ba
 \widetilde m \simeq 100-450 \mbox{ GeV}
\label{eq31n}
\ea
precisely the range where SUSY particles are often expected.  If
supersymmetry in the mass range of Eq.~\eq{eq31n} with relatively
large $\tan\beta$ is responsible for the apparent $a_\mu^{\rm
exp}-a_\mu^{\rm SM}$ difference, it will have many dramatic
consequences.  Besides expanding the known symmetries of Nature and
our fundamental notion of space-time, it will impact other new
exploratory experiments.  Indeed, for $ \widetilde m \simeq 100-450$
GeV, one can expect a plethora of new SUSY particles to be discovered
soon, either at the Fermilab 2 TeV $p\bar p$ collider or certainly at
the LHC 14 TeV $pp$ collider which is expected to start running 2006.  

Large $\tan\beta$ supersymmetry can also have other interesting
loop--induced low energy consequences beyond $a_\mu$.  For example, it
can affect $b\to s\gamma$.  Even for the muon, ``New Physics'' in
$a_\mu$ is likely to suggest potentially observable $\mu\to e\gamma$,
$\mu^-N\to e^-N$ and a muon electric dipole moment (edm), depending on
the degree of flavor mixing and $CP$ violating phases.  Searches for
these phenomena are now entering an exciting phase, with a new
generation of experiments being proposed or constructed.  The decay
$\mu\to e\gamma$ will be searched for with $2\times 10^{-14}$ single
event sensitivity (SES) at the Paul Scherrer Institute
\cite{Mori1999}.  The MECO experiment at BNL \cite{Popp:2001hu} will
search for the muon-electron conversion, $\mu^-{\rm Al}\to e^-{\rm
Al}$, with $2\times 10^{-17}$ SES.  A proposal has been made
\cite{Semertzidis:1999kv} to search for the muon's electric dipole
moment with sensitivity of about $10^{-24}$ e$\cdot$cm with the BNL
muon storage ring.  Certainly, the hint of supersymmetry suggested by
$a_\mu^{\rm exp}$ will provide strong additional motivation to extend
such studies both theoretically and experimentally.

\subsection{Radiative Muon Mass Models}
The relatively light masses of the muon and most other known
fundamental fermions could suggest that they are radiatively loop induced
by ``New Physics'' beyond the Standard Model.  Although no compelling
model exists, the concept is very attractive as a natural scenario for
explaining the flavor mass hierarchy, i.e. why most fermion masses are
so much smaller than the electroweak scale $\sim 250$ GeV. 

The basic idea is to start off with a naturally zero bare fermion mass
due to an underlying chiral symmetry.  The symmetry is broken in the
fermion 2-point function by
quantum loop effects.  They lead to a finite calculable mass which
depends on the mass scales, coupling strengths and dynamics of the
underlying symmetry breaking mechanism.  In such a scenario, 
one generically expects for
the muon 
\ba
m_\mu \propto {g^2\over 16\pi^2} M_F,
\label{eq24}
\ea
where $g$ is some new interaction coupling strength and $M_F\sim
100-1000$ GeV is a heavy scale associated with chiral symmetry
breaking and perhaps electroweak symmetry breaking.  
Of course, there may be other suppression factors at work
in Eq.~(\ref{eq24}) that keep the muon mass small.

Whatever source of chiral symmetry breaking is responsible for
generating the muon's mass will also give rise to non-Standard Model
contributions in $a_\mu$.  Indeed, fermion masses and anomalous
magnetic moments are intimately connected chiral symmetry breaking
operators.  Remarkably, in such radiative scenarios, the additional
contribution to $a_\mu$ is quite generally given by
\cite{WM:Tennessee,MassMech} 
\ba
a_\mu(\mbox{New Physics}) \simeq C{m_\mu^2 \over M^2}, \qquad C\simeq
\order{1}, 
\label{eq25}
\ea 
where $M$ is some physical high mass scale associated with the ``New
Physics'' and $C$ is a model-dependent number roughly of order 1 (it
can even be larger).  $M$ need not be the same scale as $M_F$ in
eq.~\eq{eq24}.  In fact, $M$ is usually a somewhat larger gauge or
scalar boson mass responsible for mediating the chiral symmetry
breaking interaction.  The result in eq.~\eq{eq25} is remarkably
simple in that it is largely independent of coupling strengths,
dynamics, etc.  Furthermore, rather than exhibiting the usual
$g^2/16\pi^2$ loop suppression factor, $a_\mu(\mbox{New Physics})$ is
related to $m_\mu^2/M^2$ by a (model dependent) constant, $C$, roughly
of $\order{1}$.

\begin{figure}[thb]
\hspace*{-5mm}
\begin{minipage}{16.cm}
\vspace*{-2mm}
\[
\raisebox{8mm}{$m_\mu\hspace*{3mm}\simeq$}
\hspace*{5mm}
\psfig{figure=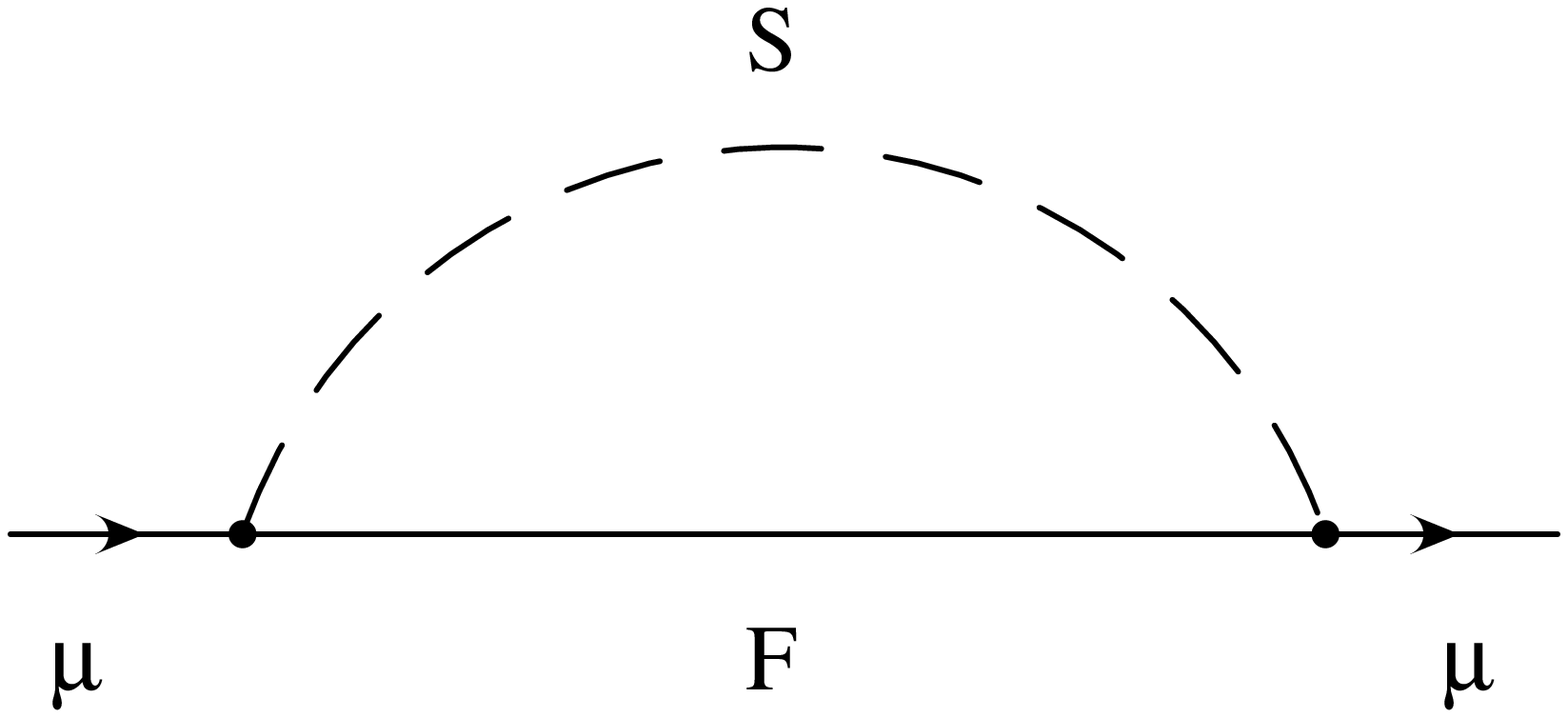,width=55mm}
\hspace*{5mm}
\raisebox{8mm}{+}
\hspace*{5mm}
\psfig{figure=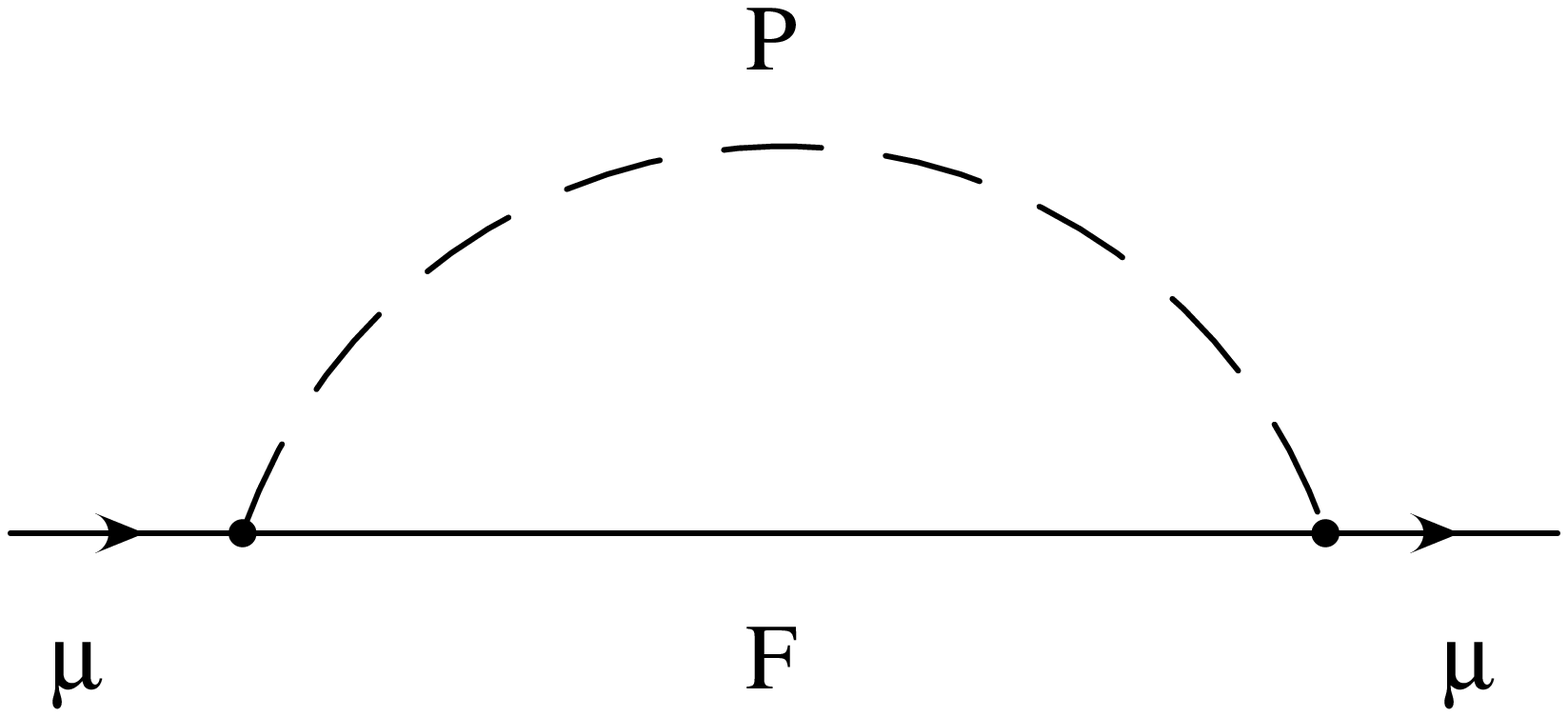,width=55mm}
\]
\end{minipage}
\caption{Example of a pair of one-loop diagrams which can induce a
finite radiative muon mass.}
\label{fig2}
\end{figure}

To demonstrate how the relationship in eq.~\eq{eq25} arises, we first
consider a simple toy model example \cite{MassMech} for muon mass
generation which is graphically depicted in Fig.~\ref{fig2}. 

If the muon is massless in lowest order (i.e.~no bare $m_\mu^0$ is
possible due to a symmetry), but couples to a heavy fermion $F$ via
scalar, $S$, and pseudoscalar, $P$, bosons with couplings $g$ and
$g\gamma_5$ respectively, then the diagrams give rise to
\ba
m_\mu &\simeq & 
 {g^2\over 16\pi^2}M_F \left(
{M_S^2 \over M_S^2 -M_F^2} \ln {M_S^2\over M_F^2}
-
{M_P^2 \over M_P^2 -M_F^2} \ln {M_P^2\over M_F^2}
\right)
\\ &\to &
{g^2\over 16\pi^2} M_F 
             \ln \left( {M_S^2\over M_P^2}\right)
\quad\mbox{($M_{S,P}\gg M_F$)}. 
\label{eq26}
\ea
Note that short-distance ultraviolet divergences have canceled and the
induced mass vanishes in the chirally symmetric limit $M_S=M_P$.  
\begin{figure}[thb]
\hspace*{-5mm}
\begin{minipage}{16.cm}
\vspace*{-2mm}
\[
\hspace*{5mm}
\psfig{figure=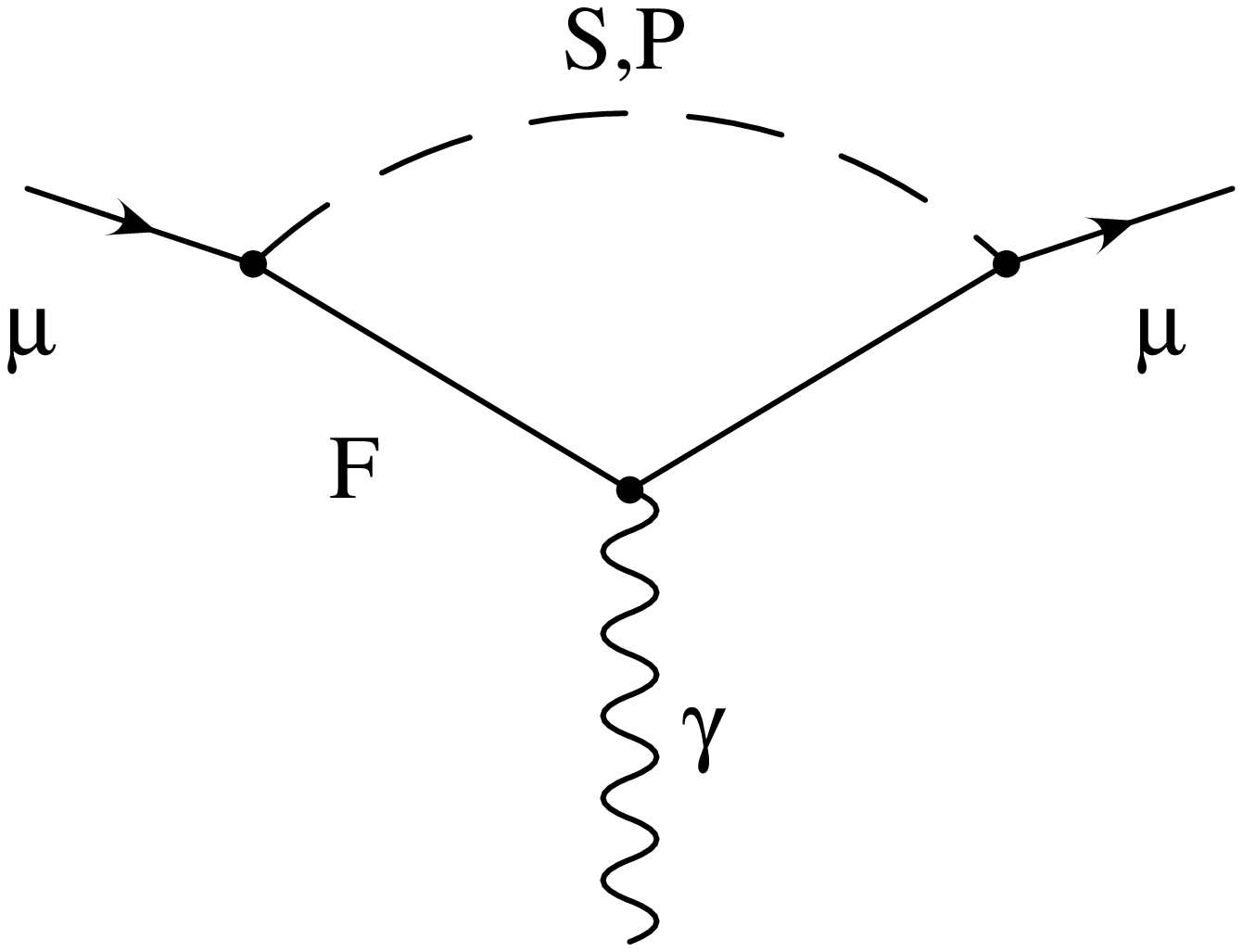,width=55mm}
\hspace*{20mm}
\psfig{figure=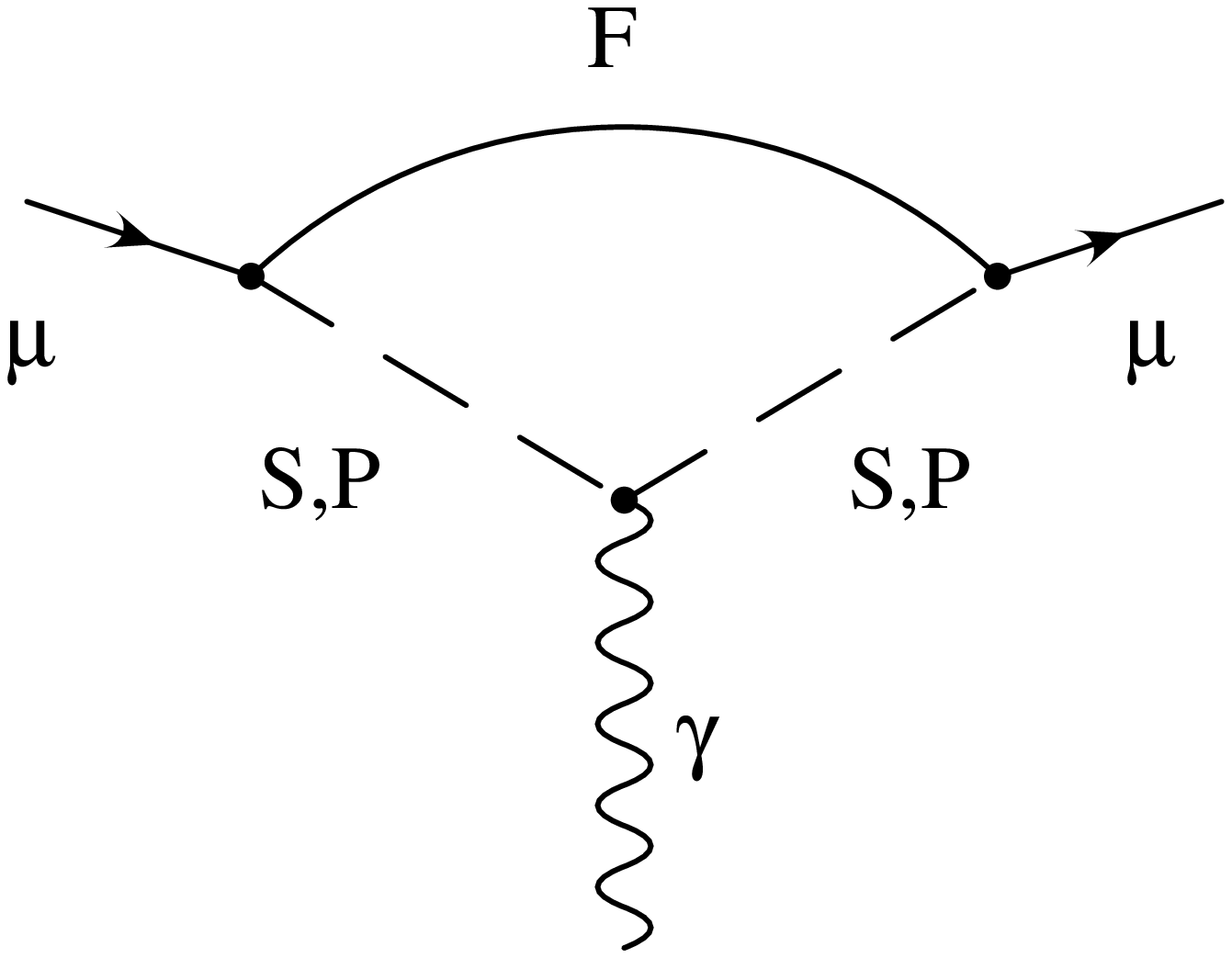,width=55mm}
\]
\end{minipage}
\caption{Potential diagrams that can contribute to the anomalous magnetic 
moment in radiative muon mass models.}
\label{fig:scal}
\end{figure}
If we attach a photon to the heavy internal fermion, $F$, or boson $S$
or $P$ (assumed to
carry fractions $Q_F$ and $1-Q_F$ of the muon charge, respectively), 
then a new contribution to $a_\mu$ is also
induced (see Fig.~\ref{fig:scal}).  
One finds 
\ba
a_\mu(\mbox{New Physics}) &=& {g^2\over 16\pi^2} \left\{
 Q_F\left[  f_N(M_P) - f_N(M_S)\right]
\right.
\nonumber \\
&&\left. \qquad
+(1-Q_F) \left[ f_C(M_P) - f_C(M_S)\right]\right\},
\ea
with
\ba
f_N(M_X) &=& {m_\mu M_F\over (M_X^2 -M_F^2)^3}
 \left( 3M_X^4 + M_F^4 -4M_X^2M_F^2 -2M_X^4 \ln{M_X^2\over
 M_F^2}\right),
\\[2mm]
f_C(M_X) &=& {m_\mu M_F\over (M_X^2 -M_F^2)^3}
\left( M_X^4-M_F^4-2M_F^2 M_X^2 \ln{M_X^2\over M_F^2}\right).
\ea
In the limit $M_{S,P}\gg M_F$ and $Q_F = 1$, one finds
\cite{MassMech}
\ba
a_\mu(\mbox{New Physics}) \simeq {g^2\over 8\pi^2}
 {m_\mu M_F \over M_P^2}
\left( {M_P^2\over M_S^2} \ln {M_S^2\over M_F^2} -  \ln {M_P^2\over
M_F^2}
\right),
\label{eq27}
\ea
while for $Q_F=0$ 
\ba
a_\mu(\mbox{New Physics}) \simeq {g^2\over 8\pi^2}
 {m_\mu M_F \over M_P^2}
\left(1-{M_P^2\over M_S^2}\right).
\label{eq39b}
\ea
The induced $a_\mu(\mbox{New Physics})$ 
also vanishes in the $M_S=M_P$ chiral symmetry limit.
Interestingly, $a_\mu(\mbox{New Physics})$ exhibits a linear rather
than quadratic dependence on $m_\mu$ at this point.  Recall, that in
section \ref{sec1} we said that such a feature was misleading or
artificial.  Our subsequent discussion should clarify that point. 

Although eqs.~\eq{eq26} and \eq{eq27} both depend on unknown
parameters such as $g$ and $M_F$, those quantities largely cancel when
we combine both expressions.  One finds
\ba
a_\mu(\mbox{New Physics}) & \simeq & C{m_\mu^2 \over M_P^2},
\nonumber \\
C&=& 2\left[ 1-\left(1-{M_P^2 \over M_S^2}\right) \ln {M_S^2 \over M_F^2}/
 \ln {M_S^2 \over M_P^2}
\right] \quad \mbox{for $Q_F=1$},
\nonumber \\
C&=& \left(1-{M_P^2 \over M_S^2}\right) / \ln {M_S^2 \over M_P^2}
\quad \mbox{for $Q_F=0$},
\ea
where $C$ is very roughly $\order{1}$.  It can actually span a broad
range and take on either sign, depending on the $M_S/M_P$ ratio and
$Q_F$.   A loop produced
$a_\mu(\mbox{New Physics})$ effect that started out at $\order{
g^2/16\pi^2}$ has effectively been promoted to $\order{1}$ by
absorbing the couplings and $M_F$ factor into $m_\mu$.  Along the way,
the linear dependence on $m_\mu$ has been replaced by a more natural
quadratic dependence.

An alternative prescription for radiatively generating fermion masses
involves new strong dynamics, e.g. extended technicolor.  In such
scenarios, technifermions acquire, via new strong dynamics, dynamical
self-energies 
\ba
\Sigma_F(p) \simeq m_F\left( \Lambda^2 \over  \Lambda^2-p^2
\right)^{1-{\gamma\over 2}},
\ea
where $0<\gamma<2$ is an anomalous dimension, $m_F\simeq \order{300
\mbox{ GeV}}$, and $\Lambda$ is the new strong interaction scale $\sim
\order{ 1\mbox{ TeV}}$.  

\begin{figure}[thb]
\hspace*{45mm}
\psfig{figure=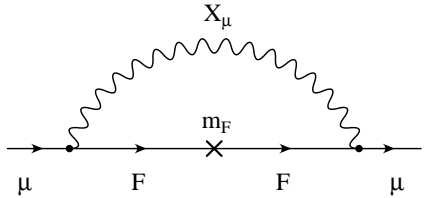,width=55mm}
\caption{Extended technicolor-like diagram responsible for generating
the  muon mass.}
\label{fig:ext}
\end{figure}

Ordinary fermions such as the muon receive loop induced masses via the
diagram in Fig.~\ref{fig:ext}.   

The extended gauge boson $X_\mu$ links $\mu$ and $F$ via the
non-chiral coupling 
\ba
g\gamma_\mu\left( a{1-\gamma_5\over 2} +
b{1+\gamma_5\over 2}\right)
\ea
and gives rise to a mass
\cite{WM:Tennessee,MassMech}
\ba
m_\mu\simeq {g^2a b\over 4\pi^2} m_F \left( {\Lambda\over m_{X_\mu}}
\right)^{2-\gamma} \Gamma\left( {\gamma\over 2}\right)
 \Gamma\left(1- {\gamma\over 2}\right),
\label{eq31new}
\ea
where $\Gamma(x)$ is the Gamma function.  Notice, the ultraviolet
divergence at $\gamma=2$ which corresponds to a non-dynamical $m_F$.

If we attach a photon to the internal fermion line in
Fig.~\ref{fig:ext} (assumed here to have charge $-1$), 
an anomalous magnetic moment contribution is
induced.   One finds
\ba
a_\mu(\mbox{New Dynamics}) \simeq
{g^2ab \over 2\pi^2} {m_\mu m_F \over m_{X_\mu}^2}
\left( {\Lambda\over  m_{X_\mu}} \right)^{2-\gamma}
{\Gamma\left(2- {\gamma\over 2}\right)
   \Gamma\left( {\gamma\over 2}\right)
 \over 1+{\gamma\over 2}}.
\label{eq32n}
\ea
Again we see only a linear dependence on $m_\mu$.  However, when
Eq.~(\ref{eq31new}) and (\ref{eq32n}) are combined, one finds
\ba
a_\mu(\mbox{New Dynamics}) \simeq 2
\left({ 2-\gamma\over 2+\gamma}\right)
{m_\mu^2 \over m_{X_\mu}^2},
\label{eq33n}
\ea
i.e. the generic result $\order{1}m_\mu^2/M^2$ where $M$ is the ``new
physics'' scale (here the extended-techniboson mass) emerges.

A similar relationship, $a_\mu(\mbox{New Physics})\simeq
Cm_\mu^2/M^2$, has been found in more realistic multi-Higgs models
\cite{Babu:1989fg},
SUSY with soft
masses \cite{Borzumati:1999sp}, etc.  It is also a natural expectation
in composite models
\cite{Brodsky:1980zm,Shaw:1980hk,Gonzalez-Garcia:1996rx} or some
models with large extra dimensions
\cite{Davoudiasl:2000my,Casadio:2000pj}, 
although studies
of
such cases have not necessarily made that same connection.  Basically,
the requirement that $m_\mu$ remain relatively small in the presence
of new chiral symmetry breaking interactions forces $a_\mu(\mbox{New
Physics})$ to effectively exhibit a quadratic $m_\mu^2$ dependence.

For models of the above variety, where $\left| a_\mu(\mbox{New
Physics}) \right| \simeq m_\mu^2/M^2$, the current constraint in
eq.~\eq{eq20} suggests (very roughly) 
\ba
M\simeq 1-2 \mbox{ TeV}. 
\label{eq47n}
\ea
Of course, for a specific model, one must check that the sign of the
induced $a_\mu^{\rm NP}$  is in accord with experiment (i.e. it should
be positive). 

 Such a scale of ``New Physics'' could be quite natural in multi-Higgs
radiative mass models and soft SUSY mass scenarios.  It would be
somewhat low for dynamical symmetry breaking, compositeness and extra
dimension models, however, confirmation of an $a_\mu^{\rm exp}$
deviation will certainly lead to all possibilities being revisited.

\subsection{Other ``New Physics'' Examples}
\subsubsection{Anomalous $W$ Boson Properties}

Anomalous $W$ boson magnetic dipole
and electric quadrupole moments can also lead to a deviation in
$a_\mu$ from SM expectations.
We generalize the $\gamma WW $ coupling such that 
the $W$ boson magnetic dipole moment is given by 
\ba
\mu_W = {e\over 2m_W} (1+\kappa+\lambda)
\ea
and electric quadrupole moment by
\ba
Q_W = -{e\over 2m_W} (\kappa - \lambda)
\ea
where $\kappa = 1$ and $\lambda = 0$ in the Standard Model,
i.e.~the gyromagnetic ratio $g_W=\kappa+1=2$. 
For non-standard couplings, one obtains the additional
one loop contribution to $a_\mu$ given by
\cite{Mery:1990dx,Herzog:1984nx,Suzuki:1985yh,Grau:1985zh,Beccaria:1998br} 
\ba
a_\mu(\kappa,\lambda) \simeq
{G_\mu m_\mu^2\over 4\sqrt{2} \pi^2}
 \left[ (\kappa-1) \ln{\Lambda^2\over m_W^2} - {1\over
3}\lambda\right],
\label{eq43n}
\ea
where $\Lambda$ is the high momentum cutoff required to give a finite
result. It presumably corresponds to the onset of ``New Physics'' such
as the $W$ compositeness scale, or new strong dynamics.  Higher order 
electroweak loop
effects reduce that contribution by roughly the suppression in
Eq.~(\ref{eq:sup}), i.e.~$\sim 9\%$. 

For $\Lambda \simeq 1$ TeV, the deviation in Eq.~\eq{eq19} corresponds
to 
\ba
\kappa-1 = 0.37\pm 0.14.
\label{eq44}
\ea
Such a large deviation from Standard Model expectations, $\kappa=1$,
is already ruled out by 
 $e^+e^-\to W^+W^-$ data at
LEP II which gives \cite{PDG2000,Przysiezniak:2000mu}
\ba 
\kappa-1 = 0.04\pm 0.08 \qquad \mbox{(LEP II)}.
\label{eq45}
\ea
One could reduce the requirement in Eq.~\eq{eq44} somewhat by assuming
a much larger $\Lambda$ cuoff in Eq.~\eq{eq43n}.  However, it is
generally felt that $\kappa-1 $ and $\Lambda$ should be inversely
correlated.  For example $\kappa-1 \sim m_W/\Lambda $ or
$(m_W/\Lambda)^2$.  So, the rather substantial $\kappa-1 $ needed to
accommodate $a_\mu^{\rm exp}$ would argue against a much larger
$\Lambda$.  Similarly, the large value of the anomalous $W$ electric
quadrupole moment $\lambda \simeq -6$ needed to reconcile $a_\mu^{\rm
exp}-a_\mu^{\rm SM}$  is also ruled out by collider data (which
implies $|\lambda| \lsim 0.1$).  Hence, it appears that anomalous $W$
boson properties cannot be the primary source of the discrepancy in
$a_\mu^{\rm exp}$.  

\subsubsection{New Gauge Bosons}
The local $SU(3)_C \times SU(2)_L \times U(1)_Y$ symmetry of the
Standard Model can be easily expanded to a larger gauge group with
additional charged and neutral gauge bosons.  Here, we consider
effects due to a charged $W^\pm_R$ which couples to right-handed
charged currents in generic left-right symmetric models and a neutral
gauge boson, $Z'$, which can naturally arise in higher rank GUT models
such as $SO(10)$ or $E_6$.  A general analysis of one-loop
contributions to $a_\mu$ from extra gauge bosons has been carried out
by Leveille  \cite{Leveille:1978rc} and the specific examples
considered here were illustrated in \cite{km90}.  Here, we will only
discuss the likelihood of such bosons being the source of the apparent 
 $a_\mu^{\rm exp}-a_\mu^{\rm SM}$ discrepancy.

For the case of a $W_R$ coupled to $\mu_R$ and a (very light) $\nu_R$
with gauge coupling $g_R$, one finds
\ba
a_\mu(W_R) \simeq (390\times 10^{-11}) {g_R^2\over g_2^2} {m_W^2 \over
m_{W_R}^2}.
\label{eq46}
\ea
To accommodate the discrepancy in Eq.~\eq{eq19} requires $m_{W_R}\simeq
m_W =80.4$ GeV  for $g_R\simeq g_2$, 
which is clearly ruled out by direct searches and
precision measurements which
give $m_{W_R}\gsim 715$ GeV.  Hence, $W^\pm_R$ is
not a viable candidate for explaining the $a_\mu^{\rm exp}$ discrepancy.

Extra neutral gauge bosons (with diagonal $\overline{\mu}\mu$
couplings) do much worse in trying to explain $a_\mu^{\rm
exp}-a_\mu^{\rm SM}$, partly 
because they often tend to give a contribution
with opposite sign.  For example, the $Z_\chi$ of $SO(10)$ leads to
\ba
a_\mu(Z_\chi) \simeq -6\times 10^{-11}
\left({m_Z^2 \over
m_{Z_\chi}^2}\right).
\label{eq47}
\ea
Given the collider constraint $m_{Z_\chi}\gsim 600$ GeV, that effect
would be much too small to observe in $a_\mu^{\rm exp}$.  Most other
$Z'$ scenarios  give similar results.

An exception to the small effects from gauge bosons illustrated above
is provided by non-chiral coupled bosons which connect $\mu$ and a
heavy fermion $F$.  In those cases, $\Delta a_\mu \simeq {g^2\over
16\pi^2} {m_\mu m_F \over M^2}$, where $M$ is the gauge boson mass.
However, loop effects then give $\delta m_\mu \sim g^2m_F$ (see the
discussion in Sect. 3.2) and we have argued that in such scenarios
$\Delta a_\mu$ should actually turn out to be $\sim m_\mu^2/M^2$.  As
previously pointed out in Eq.~\eq{eq47n}, $a_\mu^{\rm exp}-a_\mu^{\rm
SM}$ then corresponds to $M\sim 1-2$ TeV.

Many other examples of ``New Physics'' contributions to $a_\mu$ have
been considered in the literature.  A general analysis in terms of
effective interactions was presented in \cite{Escribano:1998hf}.
Specific other examples include effects due to muon compositeness
\cite{Gonzalez-Garcia:1996rx}, extra Higgs \cite{Krawczyk:1997sm}
bosons, leptoquarks \cite{Couture95,Davidson:1994qk}, bileptons
\cite{Cuypers:1996ia}, 2-loop pseudoscalar effects
\cite{Chang:2000ii}, compact extra dimensions
\cite{Graesser:1999yg,Nath:1999aa} etc. Given the apparent deviation
in experiment from theory, all will certainly be revisited.

\section{Outlook}
After many years of experimental and theoretical toil, studies of the
muon anomalous magnetic moment have entered an exciting new phase.
Experiment E821 at Brookhaven has reported a 2.6 sigma difference
betwen $a_\mu^{\rm exp}$ and the Standard Model prediction,
$a_\mu^{\rm SM}$.  That difference could be a strong hint of
supersymmetry in roughly the $\tan\beta\simeq 4$, $m_{\rm SUSY}\simeq
100$ GeV $-$ $\tan\beta\simeq 40$, $m_{\rm SUSY}\simeq 450$ GeV region
or perhaps an indication of radiative muon mass generation from ``new
physics'' in the $1-2$ TeV range.  Either case represents an exciting
prospect with interesting implications for future experiments.

Of course, before the assertion of ``new physics'' can be taken
seriously, the values of $a_\mu^{\rm exp}$ and $a_\mu^{\rm SM}$ should
be further scrutinized and refined.  In that regard, it is fortunate
that ongoing analysis of existing $\mu^+$ data should reduce the
uncertainty in $a_\mu^{\rm exp}$ by about another factor of 2.5 and
similar statistical accuracy is expected from ongoing $\mu^-$ studies.
In addition, ongoing analysis of $e^+e^-\to \pi^+\pi^-$ data in the
$\rho $ resonance region and future experimental studies at higher
energy should significantly reduce the uncertainty in $a_\mu^{\rm SM}$
and enhance its credibility.  Should a significant difference between
theory and experiment persist after these improvements, it will
rightfully be heralded as a harbinger of ``new physics''.  We look
forward to the anticipated confrontation.

\section*{Acknowledgments}
This work was supported  by the DOE under grant number
DE-AC02-98CH10886.


\begin{thebibliography}{10}

\bibitem{Dehmelt87}
R.~S. {van Dyck Jr.}, P.~B. Schwinberg, and H.~G. Dehmelt, Phys. Rev. Lett.
  {\bf 59},  26  (1987).

\bibitem{Mohr99}
P.~J. Mohr and B.~N. Taylor, Rev. Mod. Phys. {\bf 72},  351  (2000).

\bibitem{Laporta:1997zy}
S. Laporta and E. Remiddi, Acta Phys. Polon. {\bf B28},  959  (1997).

\bibitem{Hughes:1999fp}
V.~W. Hughes and T. Kinoshita, Rev. Mod. Phys. {\bf 71},  S133  (1999).

\bibitem{Czarnecki:1998nd}
A. Czarnecki and W.~J. Marciano, Nucl. Phys. B (Proc. Suppl.) {\bf 76},  245
  (1999).

\bibitem{Kinoshita:1996vz}
T. Kinoshita, Rept. Prog. Phys. {\bf 59},  1459  (1996).

\bibitem{Gab94}
G. Gabrielse and J. Tan,  in {\em Cavity Quantum Electrodynamics}, edited by
  P.~R. Berman (Academic Press, San Diego, 1994), p.\ 267.

\bibitem{PDG98}
C. {Caso \em et al. (Particle Data Group)}, Eur. Phys. J. {\bf C3},  1  (1998).

\bibitem{Brown:2000sj}
H.~N. Brown {\it et~al.}, hep-ex/0009029 (unpublished).

\bibitem{Brown2001}
H.~N. Brown {\it et~al.}, hep-ex/0102017 (unpublished).

\bibitem{km90}
T. Kinoshita and W.~J. Marciano,  in {\em Quantum Electrodynamics}, edited by
  T. Kinoshita (World Scientific, Singapore, 1990), pp.\ 419--478.

\bibitem{GRaf}
M. Gourdin and E. {de Rafael}, Nucl. Phys. {\bf B10},  667  (1969).

\bibitem{Alemany:1997tn}
R. Alemany, M. Davier, and A. H{\"o}cker, Eur. Phys. J. {\bf C2},  123  (1998).

\bibitem{Davier:1998si}
M. Davier and A. H{\"o}cker, Phys. Lett. {\bf B435},  427  (1998).

\bibitem{Davier:1999xy}
M. Davier, hep-ex/9912044 (unpublished).

\bibitem{Davier:1998iz}
M. Davier, Nucl. Phys. B (Proc. Suppl.) {\bf 76},  327  (1999).

\bibitem{Jeg95}
S. Eidelman and F. Jegerlehner, Z. Phys. {\bf C67},  585  (1995).

\bibitem{kinoshita85}
T. Kinoshita, B. Nizic, and Y. Okamoto, Phys. Rev. {\bf D31},  2108  (1985).

\bibitem{light}
E. de~Rafael, Phys. Lett. {\bf B322},  239  (1994).

\bibitem{Erler:2000nx}
J. Erler and M. Luo, hep-ph/0101010 (unpublished).

\bibitem{Marciano:1988vm}
W.~J. Marciano and A. Sirlin, Phys. Rev. Lett. {\bf 61},  1815  (1988).

\bibitem{EidelmanPriv}
S. Eidelman, private communication.

\bibitem{Jegerlehner:1999hg}
F. Jegerlehner,  in {\em Radiative Corrections}, edited by J. Sol{\`a} (World
  Scientific, Singapore, 1999), pp.\ 75--89.

\bibitem{Anderson:1999ui}
S. Anderson {\it et~al.}, Phys. Rev. {\bf D61},  112002  (2000).

\bibitem{Marciano:1992pr}
W.~J. Marciano, Phys. Rev. {\bf D45},  R721  (1992).

\bibitem{Jeg2000}
F. Jegerlehner, 2000, seminar at New York University in honor of A. Sirlin's
  70th Birthday.

\bibitem{Krause:1997rf}
B. Krause, Phys. Lett. {\bf B390},  392  (1997).

\bibitem{Bijnens:1996xf}
J. Bijnens, E. Pallante, and J. Prades, Nucl. Phys. {\bf B474},  379  (1996).

\bibitem{Hayakawa:1998rq}
M. Hayakawa and T. Kinoshita, Phys. Rev. {\bf D57},  465  (1998).

\bibitem{Brodsky:1967mv}
S.~J. Brodsky and J.~D. Sullivan, Phys. Rev. {\bf 156},  1644  (1967).

\bibitem{Burnett67}
T. Burnett and M.~J. Levine, Phys. Lett. {\bf 24B},  467  (1967).

\bibitem{Jackiw72}
R. Jackiw and S. Weinberg, Phys. Rev. {\bf D5},  2473  (1972).

\bibitem{fls72}
K. Fujikawa, B.~W. Lee, and A.~I. Sanda, Phys. Rev. {\bf D6},  2923  (1972).

\bibitem{Bars72}
I. Bars and M. Yoshimura, Phys. Rev. {\bf D6},  374  (1972).

\bibitem{ACM72}
G. Altarelli, N. Cabibbo, and L. Maiani, Phys. Lett. {\bf B40},  415  (1972).

\bibitem{Bardeen72}
W.~A. Bardeen, R. Gastmans, and B.~E. Lautrup, Nucl. Phys. {\bf B46},  315
  (1972).

\bibitem{KKSS}
T.~V. Kukhto, E.~A. Kuraev, A. Schiller, and Z.~K. Silagadze, Nucl. Phys. {\bf
  B371},  567  (1992).

\bibitem{CKM96}
A. Czarnecki, B. Krause, and W. Marciano, Phys. Rev. Lett. {\bf 76},  3267
  (1996).

\bibitem{CKM95}
A. Czarnecki, B. Krause, and W. Marciano, Phys. Rev. {\bf D52},  R2619  (1995).

\bibitem{Peris:1995bb}
S. Peris, M. Perrottet, and E. de~Rafael, Phys. Lett. {\bf B355},  523  (1995).

\bibitem{Degrassi:1998es}
G. Degrassi and G.~F. Giudice, Phys. Rev. {\bf D58},  053007  (1998).

\bibitem{fayet80}
P. Fayet,  in {\em Unification of the Fundamental Particle Interactions},
  edited by S. Ferrara, J. Ellis, and P. {van Nieuwenhuizen} (Plenum, New York,
  1980), p.\ 587.

\bibitem{Grifols:1982vx}
J.~A. Grifols and A. Mendez, Phys. Rev. {\bf D26},  1809  (1982).

\bibitem{Ellis:1982by}
J. Ellis, J. Hagelin, and D.~V. Nanopoulos, Phys. Lett. {\bf B116},  283
  (1982).

\bibitem{Barbieri:1982aj}
R. Barbieri and L. Maiani, Phys. Lett. {\bf B117},  203  (1982).

\bibitem{Romao:1985pn}
J.~C. Romao, A. Barroso, M.~C. Bento, and G.~C. Branco, Nucl. Phys. {\bf B250},
   295  (1985).

\bibitem{Kosower:1983yw}
D.~A. Kosower, L.~M. Krauss, and N. Sakai, Phys. Lett. {\bf B133},  305
  (1983).

\bibitem{Yuan:1984ww}
T.~C. Yuan, R. Arnowitt, A.~H. Chamseddine, and P. Nath, Z. Phys. {\bf C26},
  407  (1984).

\bibitem{Vendramin:1989rd}
I. Vendramin, Nuovo Cim. {\bf A101},  731  (1989).

\bibitem{Grifols:1986vr}
J.~A. Grifols, J. Sola, and A. Mendez, Phys. Rev. Lett. {\bf 57},  2348
  (1986).

\bibitem{Morris:1988fm}
D.~A. Morris, Phys. Rev. {\bf D37},  2012  (1988).

\bibitem{Frank:1988yn}
M. Frank and C.~S. Kalman, Phys. Rev. {\bf D38},  1469  (1988).

\bibitem{Francis:1991pi}
R.~M. Francis, M. Frank, and C.~S. Kalman, Phys. Rev. {\bf D43},  2369  (1991).

\bibitem{Lopez:1994vi}
J.~L. Lopez, D.~V. Nanopoulos, and X. Wang, Phys. Rev. {\bf D49},  366  (1994).

\bibitem{Moroi:1996yh}
T. Moroi, Phys. Rev. {\bf D53},  6565  (1996).

\bibitem{Cho:2000sf}
G.-C. Cho, K. Hagiwara, and M. Hayakawa, Phys. Lett. {\bf B478},  231  (2000).

\bibitem{Ibrahim:1999aj}
T. Ibrahim and P. Nath, Phys. Rev. {\bf D62},  015004  (2000).

\bibitem{Brignole:1999gf}
A. Brignole, E. Perazzi, and F. Zwirner, JHEP {\bf 09},  002  (1999).

\bibitem{Carena:1997qa}
M. Carena, G.~F. Giudice, and C.~E.~M. Wagner, Phys. Lett. {\bf B390},  234
  (1997).

\bibitem{Mahanthappa:1999ta}
K.~T. Mahanthappa and S. Oh, Phys. Rev. {\bf D62},  015012  (2000).

\bibitem{Nath95}
U. Chattopadhyay and P. Nath, Phys. Rev. {\bf D53},  1648  (1996).

\bibitem{Goto:1999mk}
T. Goto, Y. Okada, and Y. Shimizu, hep-ph/9908499 (unpublished).

\bibitem{Blazek:1999hb}
T. Blazek, hep-ph/9912460 (unpublished).

\bibitem{Chattopadhyay:2000ws}
U. Chattopadhyay, D.~K. Ghosh, and S. Roy, hep-ph/0006049 (unpublished).

\bibitem{Mori1999}
T. Mori {\it et~al.}, Search for $\mu^+\to e^+\gamma$ down to $10^{-14}$
  branching ratio, 1999, proposal to PSI. http://meg.psi.ch/doc.

\bibitem{Popp:2001hu}
J.~L. Popp {\it et~al.}, hep-ex/0101017 (unpublished).

\bibitem{Semertzidis:1999kv}
Y.~K. Semertzidis {\it et~al.}, hep-ph/0012087 (unpublished).

\bibitem{WM:Tennessee}
W.~J. Marciano,  in {\em Radiative Corrections: Status and Outlook}, edited by
  B.~F.~L. Ward (World Scientific, Singapore, 1995), pp.\ 403--414.

\bibitem{MassMech}
W. Marciano,  in {\em Particle Theory and Phenomenology}, edited by K. Lassila
  {\it et~al.} (World Scientific, Singapore, 1996), p.\ 22.

\bibitem{Babu:1989fg}
K.~S. Babu and E. Ma, Mod. Phys. Lett. {\bf A4},  1975  (1989).

\bibitem{Borzumati:1999sp}
F. Borzumati, G.~R. Farrar, N. Polonsky, and S. Thomas, Nucl. Phys. {\bf B555},
   53  (1999).

\bibitem{Brodsky:1980zm}
S.~J. Brodsky and S.~D. Drell, Phys. Rev. {\bf D22},  2236  (1980).

\bibitem{Shaw:1980hk}
G.~L. Shaw, D. Silverman, and R. Slansky, Phys. Lett. {\bf B94},  57  (1980).

\bibitem{Gonzalez-Garcia:1996rx}
M.~C. Gonzalez-Garcia and S.~F. Novaes, Phys. Lett. {\bf B389},  707  (1996).

\bibitem{Davoudiasl:2000my}
H. Davoudiasl, J.~L. Hewett, and T.~G. Rizzo, hep-ph/0006097 (unpublished).

\bibitem{Casadio:2000pj}
R. Casadio, A. Gruppuso, and G. Venturi, hep-th/0010065 (unpublished).

\bibitem{Mery:1990dx}
P. Mery, S.~E. Moubarik, M. Perrottet, and F.~M. Renard, Z. Phys. {\bf C46},
  229  (1990).

\bibitem{Herzog:1984nx}
F. Herzog, Phys. Lett. {\bf 148B},  355  (1984).

\bibitem{Suzuki:1985yh}
M. Suzuki, Phys. Lett. {\bf 153B},  289  (1985).

\bibitem{Grau:1985zh}
A. Grau and J.~A. Grifols, Phys. Lett. {\bf 154B},  283  (1985).

\bibitem{Beccaria:1998br}
M. Beccaria, F.~M. Renard, S. Spagnolo, and C. Verzegnassi, Phys. Lett. {\bf
  B448},  129  (1999).

\bibitem{PDG2000}
D.~E. {Groom \em et al. (Particle Data Group)}, Eur. Phys. J. {\bf C15},  1
  (2000).

\bibitem{Przysiezniak:2000mu}
H. Przysiezniak,  in {\em Intersections of particle and nuclear physics},
  edited by Z. Parsa and W.~J. Marciano (AIP, Melville, NY, 2000), p.\ 1.

\bibitem{Leveille:1978rc}
J.~P. Leveille, Nucl. Phys. {\bf B137},  63  (1978).

\bibitem{Escribano:1998hf}
R. Escribano and E. Masso, Eur. Phys. J. {\bf C4},  139  (1998).

\bibitem{Krawczyk:1997sm}
M. Krawczyk and J. Zochowski, Phys. Rev. {\bf D55},  6968  (1997).

\bibitem{Couture95}
G. Couture and H. K{\"o}nig, Phys. Rev. {\bf D53},  555  (1996).

\bibitem{Davidson:1994qk}
S. Davidson, D. Bailey, and B.~A. Campbell, Z. Phys. {\bf C61},  613  (1994).

\bibitem{Cuypers:1996ia}
F. Cuypers and S. Davidson, Eur. Phys. J. {\bf C2},  503  (1998).

\bibitem{Chang:2000ii}
D. Chang, W.-F. Chang, C.-H. Chou, and W.-Y. Keung, hep-ph/0009292
  (unpublished).

\bibitem{Graesser:1999yg}
M.~L. Graesser, Phys. Rev. {\bf D61},  074019  (2000).

\bibitem{Nath:1999aa}
P. Nath and M. Yamaguchi, Phys. Rev. {\bf D60},  116006  (1999).

\end{thebibliography}

\end{document}